\documentclass[a4paper]{article}

\usepackage{multicol}
\usepackage{amsthm}
\usepackage{bm}

\usepackage[dvipdfmx]{graphicx}
\usepackage{bmpsize}
\usepackage{amssymb}
\usepackage{amsmath}
\usepackage{color}
\usepackage{comment}

\definecolor{dgreen}{rgb}{0.0, 0.5, 0.0}

\begin{document}

\fontsize{14pt}{16.5pt}\selectfont

\begin{center}
\bf{Poincar\'{e} Map Method for Limit Cycles \\in a Max-Plus Dynamical System}
\end{center}
\fontsize{12pt}{11pt}\selectfont
\begin{center}
Shousuke Ohmori$^{*)}$ and Yoshihiro Yamazaki
\end{center}

\noindent
\it{Faculty of Science and Engineering, Waseda University, Shinjuku, Tokyo 169-8555, Japan}\\

\noindent
*corresponding author: 42261timemachine@ruri.waseda.jp\\
~~\\
\rm
\fontsize{11pt}{14pt}\selectfont\noindent

\baselineskip 30pt

{\bf Abstract}\\
%
Dynamical properties of limit cycles in a two-dimensional max-plus dynamical system are discussed.
We apply a Poincar\'{e} map method to the limit cycles in order to reveal their stabilities.
This method reduces the two dimensional system to 
a one-dimensional piecewise linear discrete dynamical system 
composed of the Poincar\'{e} map and its cross section.
Basins for one of the limit cycles are derived by considering the inverse system of the original model.
It is found that the obtained basins show a hierarchic structure.
Relationship between the Poincar\'{e} map method and 
the method of piecewise linear mapping studied in integrable system theory
for the limit cycles 
is discussed.
%



%
%

\section{Introduction}

An ultradiscrete equation is a difference equation with max-plus algebra.
It can be derived from a difference equation by means of ultradiscretization\cite{Tokihiro1996}.
This method converts a difference equation into ultradiscretized one by the following two procedures;
(i) a positive variable in a difference equation, say $x_n$,  is transformed into $X_n$ by $x_n=\exp (X_n/\varepsilon)$,
where $\varepsilon$ is a positive parameter, 
and (ii) the ultradiscrete limits
	\begin{eqnarray}
\begin{cases}
		\displaystyle\lim_{\varepsilon  \to +0} \varepsilon  \log(e^{A/\varepsilon }+e^{B/\varepsilon }+\cdot \cdot \cdot )~=~\max(A,B,\cdots),\\
		\displaystyle\lim_{\varepsilon  \to +0} \varepsilon  \log(e^{A/\varepsilon }\cdot e^{B/\varepsilon }\cdot \dots ) = A+B+ \cdots ,
\end{cases}
		\label{eqn:0}
	\end{eqnarray}
are imposed.
The ultradiscretization method has been successfully applied to integrable systems\cite{Grammaticos1997,Takahashi1997}.
%
%
One typical example is ultradiscrete Burgers equation\cite{Nishinari1998}.

The ultradiscretization method has also been applied to non-equilibrium dissipative systems 
such as reaction-diffusion systems
\cite{Murata2013,Ohmori2014,Matsuya2015,Murata2015,Ohmori2016,Ohmori2020,Yamazaki2021,Ohmori2021,Isojima2022}.
Especially, we have focused on application to dynamical systems\cite{Ohmori2020,Yamazaki2021,Ohmori2021}. 
In general, ultradiscrete equations are composed of piecewise linear difference equations,
and their dynamical properties, e.g., existence of fixed points and their stabilities, 
can be characterized as discrete dynamical system.  
In our studies, the dynamical properties of the ultradiscrete equations 
that exhibit bifurcations, namely, ultradiscrete bifurcations, have been discussed.

Here, we consider the following set of max-plus equations \cite{Yamazaki2021,Ohmori2021}: 
\begin{eqnarray}
	X_{n+1} & = & Y_n + \max(0,2X_n),
	\label{eqn:1-1a} \\
	Y_{n+1} & = & B-\max(0,2X_n).
	\label{eqn:1-1b}
\end{eqnarray} 
These equations are obtained from Sel'kov model via tropical discretization and ultradiscretization.
We found that this model has the two different limit cycles, i.e., periodic solutions around an unstable fixed point.
These limit cycles possess the following characteristic dynamical properties;
(i) they are composed of seven states.
(ii) any initial states without the unstable fixed point converges to one of the two limit cycles with finite iteration time steps.

In this paper, dynamical properties of these limit cycles are analytically discussed.
By using Poincar\'{e} map, we rigorously characterize difference between the two limit cycles.
In the next section, we review the dynamical properties of eqs.(\ref{eqn:1-1a})-(\ref{eqn:1-1b}) with $B>0$. 
By introducing a Poincar\'{e} section for the limit cycles and its Poincar\'{e} map,
stabilities of these cycles are revealed in Sec.3.
In Sec. 4, basins for one of the limit cycles are clarified.
The discussion and the conclusion are given in Sec. 5 and 6, respectively.

\section{Review for the limit cycles obtained from Eqs. (\ref{eqn:1-1a})-(\ref{eqn:1-1b})}
\label{sec.2}

To investigate the dynamical properties of eqs.(\ref{eqn:1-1a})-(\ref{eqn:1-1b}), 
it is convenient to consider the two regions in $X_nY_n$-plane: 
I $\{(X_n,Y_n) ; X_n>0\}$ and II $\{(X_n,Y_n) ; X_n\leq 0 \}$. 
Equations (\ref{eqn:1-1a})-(\ref{eqn:1-1b}) can be rewritten by the following matrix form: 
\begin{eqnarray}
		\left(
   			\begin{array}{ccc}
      		X_{n+1}  \\
      		Y_{n+1}  
   			\end{array}
  		\right)
		= 
		\bm{A}
		\left(
    		\begin{array}{ccc}
      		X_{n}  \\
      		Y_{n}  
    		\end{array}
  		\right)
		+
		\left(
    		\begin{array}{ccc}
      		0  \\
      		B  
    		\end{array}
  		\right),
		\label{eqn:2-1}
	\end{eqnarray} 
where 
\begin{eqnarray}
	\bm A	=
		\begin{cases}
			\bm A_I = \left(
   			\begin{array}{ccc}
      		2 & 1  \\
      		-2 & 0  \\
   			\end{array}
  		\right) \;\;\;\; \mbox{when $(X_n,Y_n)$ is in region I}, 
		\\
		\bm A_{II} = \left(
   			\begin{array}{ccc}
      		0 & 1  \\
      		0 & 0  \\
   			\end{array}
  		\right) \;\;\;\; \mbox{when $(X_n,Y_n)$ is in region II}.
		\label{eqn:2-2}
	\end{cases}
\end{eqnarray}
From $\bm A_I$, it is found that the solution trajectory becomes a clockwise spiral 
moving away from the unstable fixed point $(B,-B)$.
For $\bm A_{II}$, $(B,B)$ is a stable fixed point 
and any $\bm{x}_0=(X_0,Y_0)$ in region II has the subsequent trajectory composed of the three points 
$\{\bm{x}_0=(X_0,Y_0), \bm{x}_{1}=(Y_0,B), \bm{x}_{2}=(B,B)\}$.

\begin{figure}[h!]
\begin{center}
\includegraphics[width=6cm]{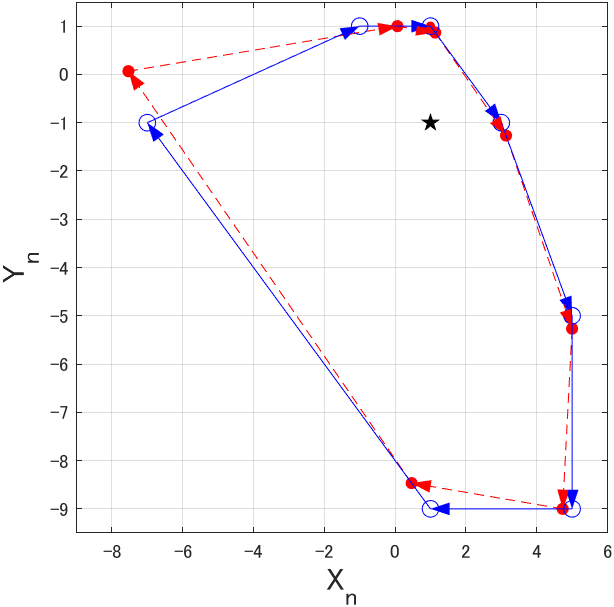}
\caption{\label{Fig.0} 
	The trajectories of $\mathcal {C}$ (open circles) and $\mathcal {C}_s$ (solid circles) for $B=1$\cite{Ohmori2021}.
	The black star shows $(B,-B)=(1,-1)$.
			}
\end{center}
\end{figure}

Hereafter we set $B>0$.
Note that both $(B,\pm B)$ are in region I.
Previously we found the two clockwise periodic solutions $\mathcal{C}$ and $\mathcal{C}_s$\cite{Ohmori2021};
each of them is composed of the seven points shown in Table \ref{Tbl.1} and Fig. \ref{Fig.0}.
%
%
%
We have also numerically shown that $\mathcal{C}$ and $\mathcal{C}_s$ are limit cycles.
In the next section, we prove this fact rigorously by means of a Poincar\'{e} map 
defined on a Poincar\'{e} section for the limit cycles.

\begin{table}[h]
   \caption{
    The seven states in limit cycles $\mathcal{C}$ and $\mathcal{C}_s$.
	 We denote these seven points in $\mathcal{C}$ and $\mathcal{C}_s$
	 as $\{ \bm{x}_j^{\mathcal{C}} \}$ and $\{ \bm{x}_j^{\mathcal{C}_s} \}$, respectively,
	 where $j=0,\cdots ,6$.
	 }

\noindent
	\begin{center}
	\begin{tabular}{|c|c|c|}
	\hline
	 & $\mathcal{C}$ & $\mathcal{C}_s$ \\
	\hline
	$\bm{x}_0$	& $(B,B)$ & $(\frac{B}{15},B)$ \\
%
	$\bm{x}_1$	& $(3B,-B)$ & $(\frac{17B}{15},\frac{13B}{15})$ \\
%
	$\bm{x}_2$	& $(5B,-5B)$ & $(\frac{47B}{15},-\frac{19B}{15})$ \\
%
	$\bm{x}_3$	& $(5B,-9B)$ & $(5B,-\frac{79B}{15})$ \\
%
	$\bm{x}_4$	& $(B,-9B)$ & $(\frac{71B}{15},-9B)$ \\
%
	$\bm{x}_5$	& $(-7B,-B)$ & $(\frac{7B}{15},-\frac{127B}{15})$ \\
%
	$\bm{x}_6$	& $(-B,B)$ & $(-\frac{113B}{15},\frac{B}{15})$ \\
%
%
    \hline
\end{tabular}
\end{center}
\label{Tbl.1}
\end{table}

%


\section{Dynamical properties of  $\mathcal{C}$ and $\mathcal{C}_s$}
\label{sec.3}

\begin{figure}[h!]
\begin{center}
\includegraphics[width=8.5cm]{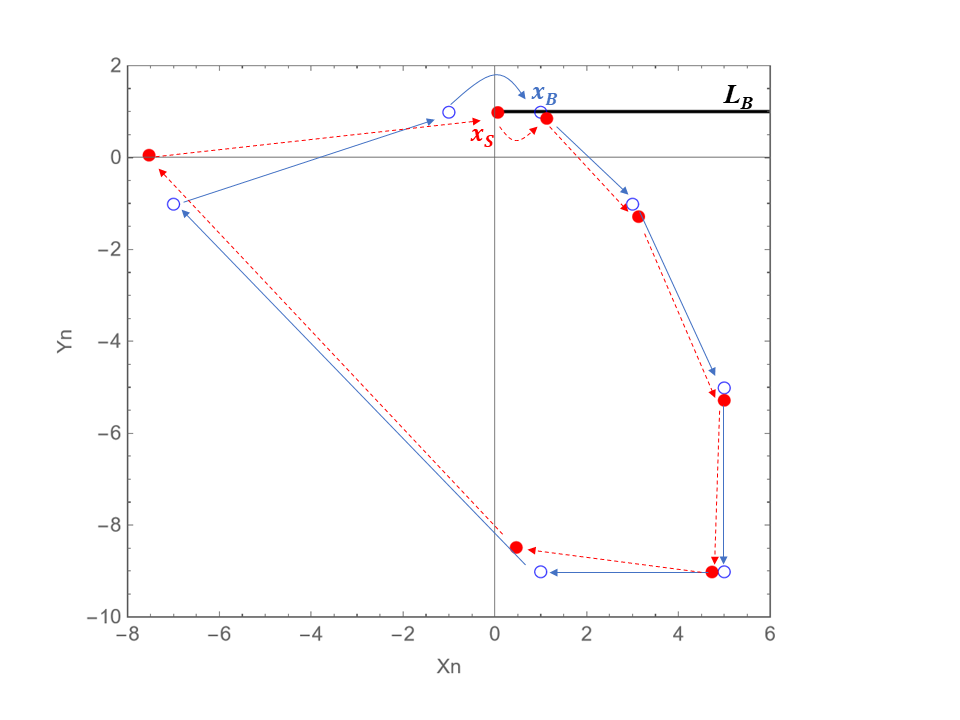}
\caption{\label{Fig.3-1} 
	Definition of the Poincar\'{e} section $L_B$ for $\mathcal {C}$ (open circle) and $\mathcal {C}_s$ (filled circle) with $B=1$.
			}
\end{center}
\end{figure}

Equation (\ref{eqn:2-1}) shows that a point in region I moves away from the fixed point clockwise, 
i.e., the fixed point becomes a unstable focus, and reaches region II after some iteration steps.
%
%
When $\bm{x}_{0}=(X_0,Y_0)$ belongs to region II, 
 then $\bm{x}_{1}=(Y_0,B)$. 
%
%
Therefore, every trajectory must pass a line 
$L_{B} \equiv \{\bm{x}_n = (X_n,B),X_n\in (0, \infty)\}$.
In particular, $\bm{x}_{0}^{\mathcal{C}}$ and $\bm{x}_{0}^{\mathcal{C}_s}$ 
in Table I return to themselves on $L_{B}$ as shown in Fig. \ref{Fig.3-1}.
Therefore, $L_B$ can be regarded as  Poincar\'{e} section for $\mathcal {C}$ and $\mathcal {C}_s$.
Next, we focus on trajectories starting at a point on $L_{B}$ and construct a Poincar\'{e} map.

\subsection{Construction of the Poincar\'{e} Map}
\label{sec.3-1}

Set $\bm{x}_0=(X_0,B)$ on $L_B$, where $X_0>0$.
Based on eq.(\ref{eqn:2-1}) with $\bm A	= \bm A_I$, 
the trajectory starting from $\bm{x}_0$ becomes as follows:
\begin{eqnarray}
	\{\bm{x}_{0} & = & (X_0,B) 
	\rightarrow \bm{x}_{1}=(2X_0+B,-2X_0+B) 
	\rightarrow \bm{x}_{2}=(2X_0+3B,-4X_0-B) \nonumber\\
	\rightarrow \bm{x}_{3} & = & (5B,-4X_0-5B) 
	\rightarrow \bm{x}_{4}=(-4X_0+5B,-9B) 
	\rightarrow \bm{x}_{5}=(-8X_0+B,8X_0-9B) \nonumber\\
	\rightarrow \bm{x}_{6} & = & (-8X_0-7B,16X_0-B)
	\rightarrow \bm{x}_{7}  =  (-B+16X_0, B) 
	\rightarrow \cdots \}.
	\label{eqn:3-1}
\end{eqnarray}
To obtain the first return points for the trajectory for eq.(\ref{eqn:2-1}) on $L_B$,
which allow us the construction of Poincare map,
we consider the following cases depending on the sign of $x$-component of $\bm{x}_{4} \sim \bm{x}_{7}$ in  (\ref{eqn:3-1}).

\begin{description}
	\item[Case 1.] When $\frac{5B}{4} \leq X_0$,
		$\bm{x}_{4}$ is in region II-1, $X_0 \leq 0, Y_0 \leq 0$. 
		Then $\bm{x}_{0}=(X_0,B)$ reaches $\bm{x}_{0}^{\mathcal{C}}=(B,B)$ with six iteration step.
%
%
%
	\item[Case 2.] When $\frac{9B}{8}<X_0<\frac{5B}{4}$, 
		$\bm{x}_0 \sim \bm{x}_4$ in region I, $\bm{x}_5$ is in region II-2, $X_0 \leq 0, Y_0 >0$, since $-8X_0+B<0$ and $0<8X_0-9B$.
		Then, $\bm{x}_6 = (8X_0-9B, B)$ becomes on $L_B$ where $0 < 8X_0-9B < B$.
%
%
	\item[Case 3.] When $\frac{B}{8}<X_0 \leq \frac{9B}{8}$, 	$\bm{x}_5$ is in region II-1, since $-8X_0+B<0$ and $8X_0-9B\leq 0$.
		Then $\bm{x}_{0}=(X_0,B)$ reaches $\bm{x}_{0}^{\mathcal{C}}$ with seven iteration steps.
%
%
%
	\item[Case 4.] When $\frac{B}{16}<X_0<\frac{B}{8}$, $\bm{x}_0 \sim \bm{x}_5$ in region I.
		$\bm{x}_6$ is in region II-2, since $16X_0-B>0$.
		Then we have $\bm{x}_{7}=(-B+16X_0,B)$ where $0<-B+16X_0<B$ on $L_B$.
	\item[Case 5.] When $X_0 \leq \frac{B}{16}$, $\bm{x}_6$ is in region II-1, 
					and hence $\bm{x}_{0}=(X_0,B)$ reaches $\bm{x}_{0}^{\mathcal{C}}$ with eight iteration steps.
\end{description}

By considering these cases, we can construct Poincar\'{e} map, which is an iteration map, $P : (0,\infty) \to (0,B]$, as follows;
 	\begin{eqnarray}
	P(X_0)=	\begin{cases}
				 B   &  (0<X_0 \leq \frac{B}{16})  \\
				16X_0-B  &  (\frac{B}{16}<X_0 < \frac{B}{8})  \\
				B & (\frac{B}{8} \leq X_0 \leq \frac{9B}{8}) \\
				8X_0-9B & (\frac{9B}{8} < X_0 < \frac{5B}{4}) \\
				B  & (\frac{5B}{4} \leq X_0).
			\label{eqn:3-1-3}
		\end{cases}
	\end{eqnarray}
%
%
In general, the discrete dynamical system composed of Poincar\'{e} map and its section can 
characterize dynamical properties of periodic solutions such as $\mathcal {C}$ and $\mathcal {C}_s$\cite{Guckenheimer,Robinson}. 
%


\subsection{Graphical analysis of $(P,(0,\infty))$}
\label{sec.3-2}

\begin{figure}[h!]
\begin{center}
\includegraphics[width=9cm]{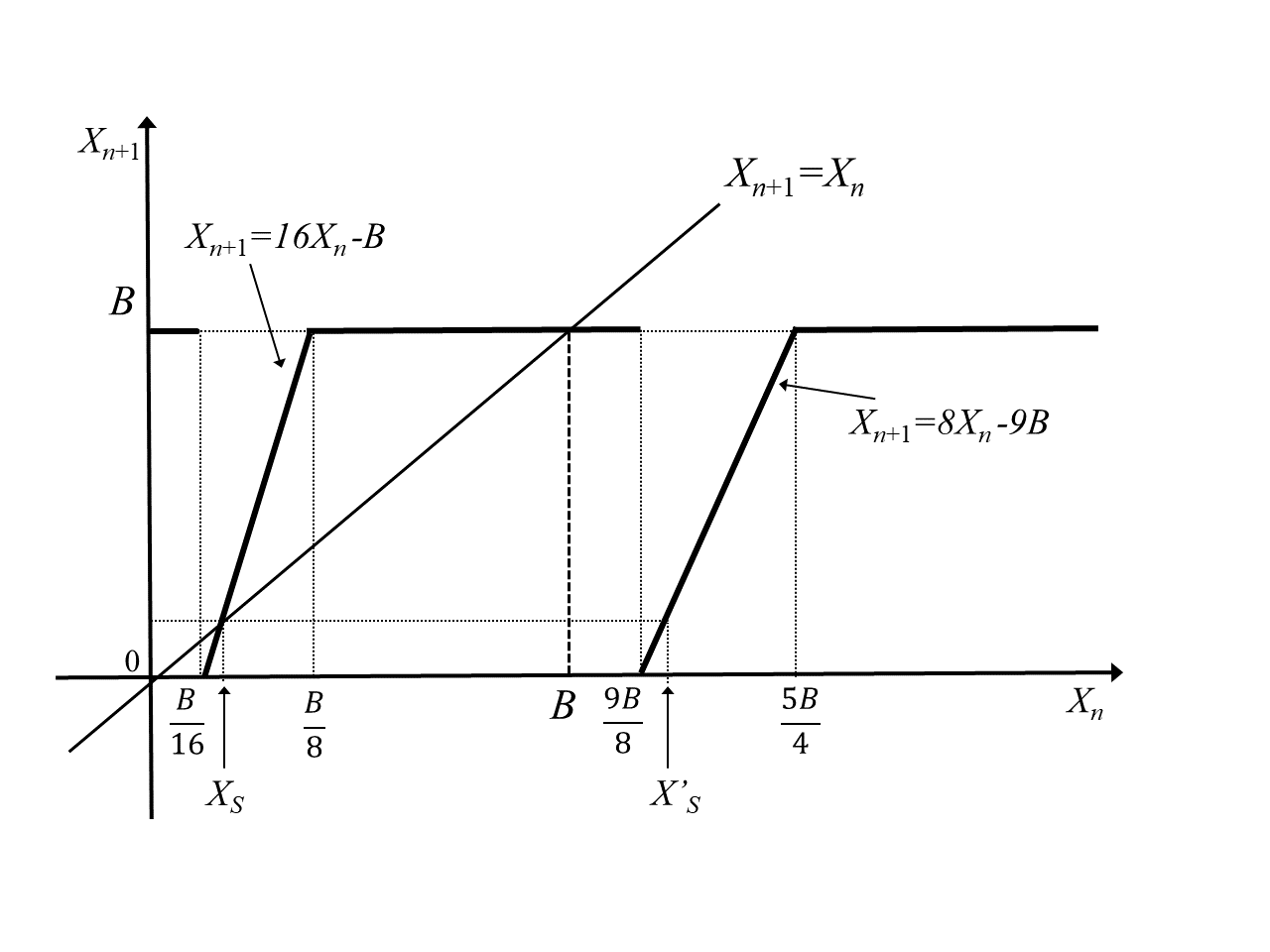}
\\
\caption{\label{Fig.P_map} 
	The graphs of the Poincar\'{e} map $P(X)$, eq.(\ref{eqn:3-1-3}).
			}
\end{center}
\end{figure}

We investigate the dynamical properties of $P$ in detail.
Equation (\ref{eqn:3-1-3}) forms a one-dimensional piecewise linear discrete dynamical system.
This dynamical properties  
can be easily grasped by using the graphical analysis\cite{Robinson}.
Figure \ref{Fig.P_map} shows the graph of $X_{n+1}=P(X_n)$.
The graph intersects the diagonal $X_{n+1}=X_{n}$ at two points $X_n=X_s\equiv \frac{B}{15}$ and $X_n=B$,
which are fixed points.
These dynamical properties obtained by graphical analysis can be summarized as follows;

\begin{description}
	\item[(i)] 
		When $0<X_0\leq \frac{B}{16}$, $\frac{B}{8} \leq X_0 \leq \frac{9B}{8}$, or $\frac{5B}{4} \leq X_0$, $P(X_0)=B$.
	\item[(ii)]
		When $\frac{B}{16}<X_0<X_s$ $(X_s<X_0<\frac{B}{8})$, there exists an iteration step $m$, 
		where $0<P^m(X_0)\leq \frac{B}{16}$ $(\frac{B}{8} \leq P^m(X_0)<B)$ and $P^{m+1}(X_0)=B$.
	\item[(iii)] 
		When $\frac{9B}{16}<X_0<\frac{5B}{4}$, $0<P(X_0)=8X_0-9B<B$.
		Especially when $X_0=X_s' \equiv \frac{17B}{15}$, $P(X_s')=X_s$ otherwise there exists $m$ where $P^{m}(X_0)=B$.
\end{description}
Note that $X=B$ is a stable fixed point, whereas $X=X_s$ is a unstable fixed point.
Therefore, we can conclude that $\mathcal {C}$ is a stable limit cycle and $\mathcal {C}_s$ is an unstable one.
%
%
%

%
%

\section{Basins of $\mathcal{C}_s$}
\label{sec.4}

In the previous section, we show that $\mathcal {C}_s$ is the unstable limit cycle.
However, the property (iii) suggests that there are some regions in $X_nY_n$-plane whose points converge to $\mathcal {C}_s$.
Here, we find such regions, namely, basins of $\mathcal{C}_s$.

\subsection{Basins in region II}
\label{sec.4-1}

\begin{figure}[h!]
\begin{center}
\includegraphics[width=7cm]{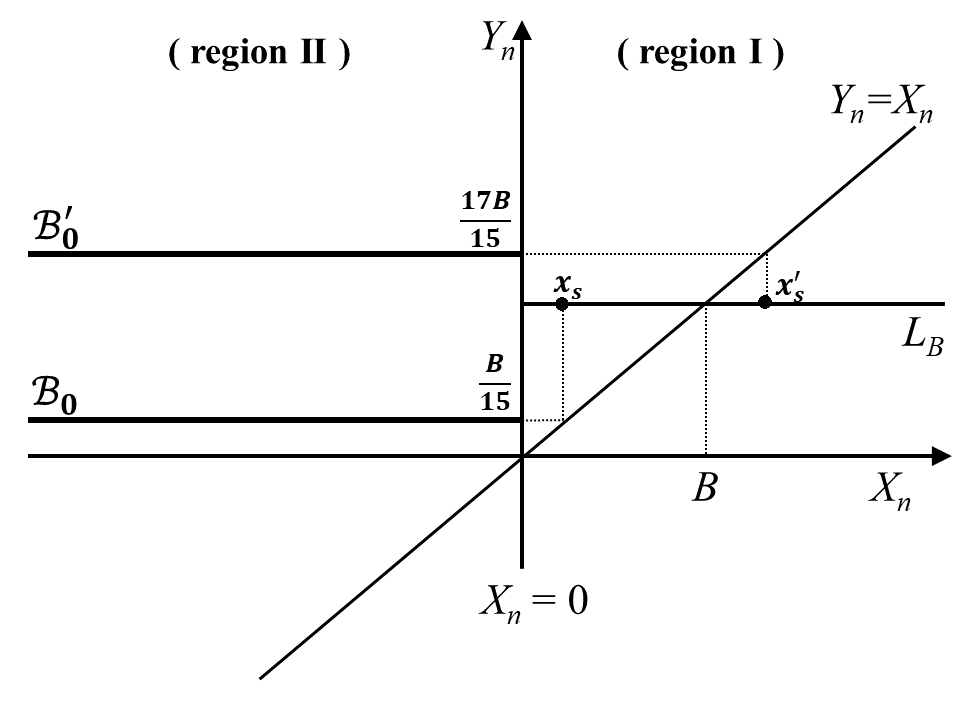}
\caption{\label{Fig.4-1} 
	Basins of $\mathcal{C}_s$ in region II.
			}
\end{center}
\end{figure}

As shown in the previous section, 
only two points $\bm{x}_{s}=(\frac{B}{15},B)$ and $\bm{x}_{s}'=(\frac{17B}{15},B)$ on $L_B$ reach $\mathcal{C}_s$.
Then, we can find the following two basins $\mathcal{B}_0$ and $\mathcal{B}_0'$ in region II-2, shown in Fig. \ref{Fig.4-1}, 
whose points arrive at $\bm{x}_{s}$ or $\bm{x}_{s}'$ with next iteration step:
\begin{eqnarray}
	\mathcal{B}_0=\{(X_0,\frac{B}{15}); X_0 \leq 0\},~\mathcal{B}_0'=\{(X_0,\frac{17B}{15}); X_0 \leq 0\}.
	\label{eqn:BI-1}
\end{eqnarray} 
Note that there is no points in region II-1 which becomes a basin for $\mathcal{C}_s$.

\subsection{Basins in region I}
\label{sec.4-2}

To elucidate basins of $\mathcal{C}_s$ in region I, 
the following backward dynamical system 
for eq.(\ref{eqn:2-1}) with $\bm A	= \bm A_I$ is considered:
\begin{eqnarray}
		\left(
   			\begin{array}{ccc}
      		X_{n+1}  \\
      		Y_{n+1}  
   			\end{array}
  		\right)
		= 
		\bm A_I^{-1}
		\left(
    		\begin{array}{ccc}
      		X_{n}  \\
      		Y_{n}  
    		\end{array}
  		\right)
		+
		\left(
    		\begin{array}{ccc}
      		\frac{B}{2}  \\
      		-B  
    		\end{array}
  		\right) \;\; ,
		\label{eqn:4-2}
	\end{eqnarray} 
where 
%
	$\bm A_I^{-1}	= \left(
		\begin{array}{ccc}
      		0 & -\frac{1}{2}  \\
      		1 & 1  \\
   			\end{array}
  		\right). $
%
%
We denote the backward trajectory of $\bm{x}_{0}$ 
obtained from eq.(\ref{eqn:4-2}) 
as $\{\bm{x}_{0}, \bm{x}^{inv}_{1}, \bm{x}^{inv}_{2}, \bm{x}^{inv}_{3}, \cdots\}$ $(=\{\bm{x}^{inv}_{n}\})$, hereafter.
Clearly the relation 
\begin{eqnarray}
	\bm{x}^{inv}_{i+1}=\bm A_I^{-1}  \bm{x}^{inv}_{i}+(\frac{B}{2},-B) 
	\Leftrightarrow \bm{x}_{i}^{inv}=\bm A_I  \bm{x}^{inv}_{i+1}+(0,B)
	\label{eqn:4-4}
\end{eqnarray}
holds for $i=0,1,\cdots,$ $(\bm{x}^{inv}_0=\bm{x}_0)$.
%
%
%
%
If $(X_0,Y_0)$ is in region I and its next iteration point is in $\mathcal{B}_0'$, 
then $(X_1,\frac{17B}{15})=\bm A_I  (X_0,Y_0)+(0,B)$ is satisfied 
and we obtain $X_0=-\frac{B}{15}<0$.
%
%
Focusing on the trajectory with a point in both region I and the basin of $\mathcal{C}_s$, 
any point in this trajectory of region II belongs only to $\mathcal{B}_0$.
Therefore, there is no point in region I whose trajectory has a point on $\mathcal{B}_0'$.

Here we set the initial point $\bm{x}_{0}$ in $\mathcal{B}_0$, 
namely $\bm{x}_{0}(z)=(z, \frac{B}{15})$, where $z \leq 0$.
Based on eq.(\ref{eqn:4-2}), $\bm{x}_{0}(z)$ becomes $\bm{x}^{inv}_{n}(z)=(X_{n}^{inv}(z), Y_{n}^{inv}(z))$, where 
\begin{eqnarray}
	X_{n}^{inv}(z) & = & B+2^{-\frac{n}{2}}\{\cos(\frac{\pi}{4}n)(z-B)-\sin(\frac{\pi}{4}n)(z+\frac{B}{15})\}, \nonumber \\
	Y_{n}^{inv}(z) & = & -2X_{n+1}^{inv}(z)+B.
	\label{eqn:4-3}
\end{eqnarray}
%
%
Now we consider a region of $z$ in which $X_{n}^{inv}(z)>0$.
Note that the term $\{\cos(\frac{\pi}{4}n)(z-B)-\sin(\frac{\pi}{4}n)(z+\frac{B}{15})\}$ in eq.(\ref{eqn:4-3}) 
has the same value for $n = \ell + 8m$ 
$(\ell = 1, 2, \cdots, 8$, and $m=0,1,2,\cdots)$.
When $\ell = 8$, $n=8(m+1)$, 
eq.(\ref{eqn:4-3}) becomes $X_{n}^{inv}(z)=2^{-\frac{n}{2}}z+(1-2^{-\frac{n}{2}})B$.
Then, $z>(1-2^{\frac{n}{2}})B$ brings about $X_{n}^{inv}(z)>0$.
According to the relation $(1-2^{\frac{8m}{2}})B>(1-2^{\frac{8(m+1)}{2}})B$, 
we have $X_{n}^{inv}(z)>0$ for $n = 8m$ if $z>(1-2^{\frac{8}{2}})B=-15B$.
Namely, whenever $z>-15B$, $\bm{x}^{inv}_{n}(z)$ always belongs to region I.
Applying this consideration to $n=\ell+8m,\ell=1\sim7, m=0,1,2,\cdots$, 
the sufficient conditions for $z$ that $\bm{x}_{n}^{inv}(z)$ belongs to region I can be obtained as follows; 
\begin{eqnarray}
	\mbox{(i)}   & z\leq 0 & \mbox{for } n=\ell+8m, \; \ell=1 \sim 5, \nonumber\\ 
	\mbox{(ii)}  & \displaystyle -\frac{121}{15}B \leq z \leq 0 & \mbox{for } n=6+8m, \nonumber\\ 
	\mbox{(iii)} & \displaystyle -\frac{113}{15}B \leq z \leq 0 & \mbox{for } n=7+8m, \nonumber\\ 
	\mbox{(iv)}  & \displaystyle -15B < z \leq 0 & \mbox{for } n=8(m+1). \nonumber
\end{eqnarray}
The condition (i) shows that $\bm{x}_{1}^{inv}(z)$ with $z\leq 0$ is in region I 
and reaches $\mathcal{B}_0$ with next iteration step 
by eq.(\ref{eqn:2-1}).
Therefore, $\mathcal{B}_1 = \{\bm{x}_{1}^{inv}(z), z\leq 0\}$ 
is a basin of $\mathcal{C}_s$ in region I.
Similarly, we can obtain the other basins in region I 
by considering (i) $\sim$ (iv), as follows:
\begin{eqnarray}
	\mathcal{B}_n & = & \{\bm{x}_{n}^{inv}(z), z\leq 0\} 
								\;\;\;\; \mbox{for  } n=1 \sim 5, \\
	\label{eqn:BII-1}
	\mathcal{B}_6 & = & \{\bm{x}_{6}^{inv}(z), -\frac{121B}{15} \leq z\leq 0\}, \\
	\label{eqn:BII-2}
	\mathcal{B}_n & = & \{\bm{x}_{n}^{inv}(z), -\frac{113B}{15} \leq z\leq 0\} 
								\;\;\;\; \mbox{for  } 7\leq n.
	\label{eqn:BII-3}
\end{eqnarray}
It is noted that when $\bm{x}_{0}$ is in a basin $\mathcal{B}_n$, 
the trajectory $\{\bm{x}_{n}\}$ passes successively all basins $\mathcal{B}_j$ with $j=n,n-1,\cdots,1$, and finally reach $\mathcal{B}_0$.
%
%

\subsection{Properties of Basins}
\label{sec.4-3}

\begin{figure}[h!]
\begin{center}
\includegraphics[width=9cm]{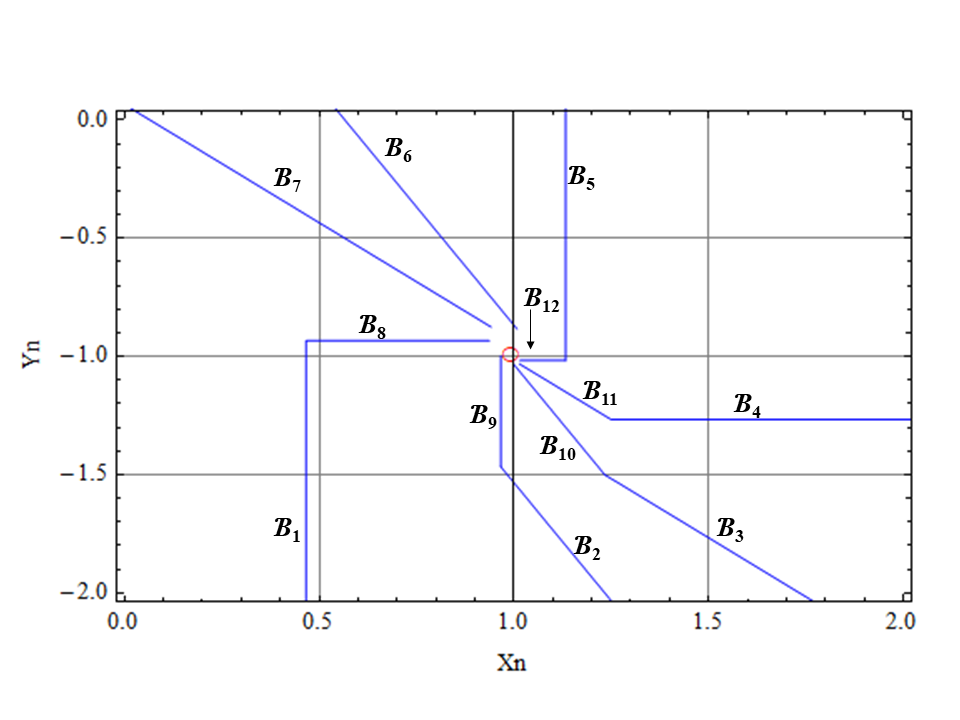}
\caption{\label{Fig.connection} 
Basins $\mathcal{B}_1 \sim \mathcal{B}_{12}$. 
These are distributed around an unstable fixed point $(B,-B)$ depicted by red circle.  
			}
\end{center}
\end{figure}

\begin{figure}[h!]
\begin{center}
\includegraphics[width=6cm]{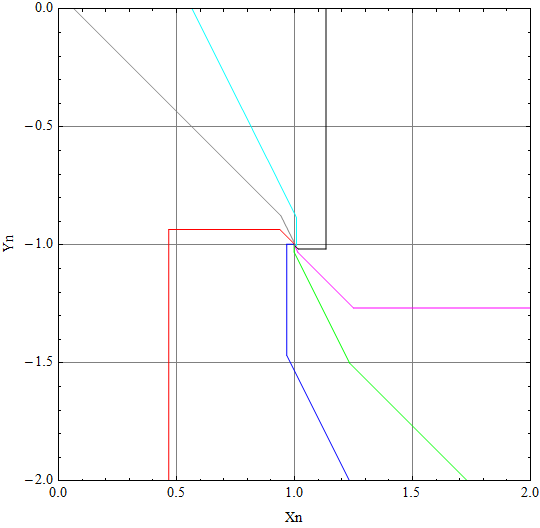}
\includegraphics[width=6cm]{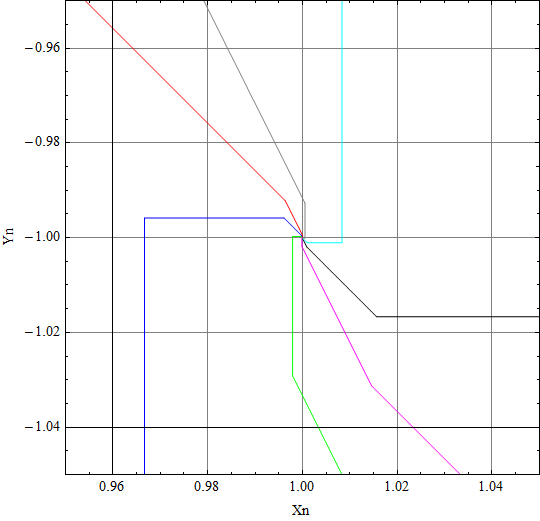}
\\
(a)
\hspace{8cm}
(b)\\
\caption{\label{Fig.self-similar} 
	Self-similarity of basins in region I.
	Both figures show basins $\mathcal{B}_1 \sim \mathcal{B}_{50}$ with $B=1$ in the region 
	(a) $[0,2]\times [-2,0]$, 
	(b) $[0.95, 1.05]\times [-1.05, -0.95]$.
	The basins' colors (red, blue, green, magenta, black, cyan, and gray) represent their connection; 
	the basins which have the same color are connected each others.
			}
\end{center}
\end{figure}

We comment some properties of the basins $\mathcal{B}_n$, $n=0,1,2,\cdots$, and $\mathcal{B}'_0$.
(i) $\mathcal{B}_6$ and $\mathcal{B}_5$ include 
$\bm{x}_{s}=(\frac{B}{15},B)$ and $\bm{x}_{s}'=(\frac{17B}{15},B)$ on $L_B$, respectively.
%
%
(ii) each $\bm{x}_{j}^{\mathcal{C}_s}$ $(j=1\sim 6)$ coincides with $\bm{x}_{6-j}^{inv}(-\frac{113B}{15})$ on $\mathcal{B}_{6-j}$.
(iii) $\bm{x}_{0}$ in $\mathcal{B}_n$ has 
the trajectory $\{\bm{x}_{0}, \bm{x}_{1}, \bm{x}_{2}, \cdots, \bm{x}_{n}\}$, 
each $\bm{x}_{j}$ passing successively $\mathcal{B}_{n-j}$ with $j=0 \sim n$, 
and then $\bm{x}_{n+1}=\bm{x}_s =\bm{x}_0^{\mathcal{C}_s}$.
(iv) For $n=0,1,2,\cdots$, $\mathcal{B}_n$ connects with $\mathcal{B}_{n+7}$ as shown in Fig. \ref{Fig.connection}. 
Indeed, the following relation holds:
\begin{eqnarray}
	\bm{x}^{inv}_{n}(0) = \bm{x}^{inv}_{n+7}(-\frac{113B}{15}) \;\;\;\; \mbox{for  } n=0,1,2\cdots .
	\label{eqn:4-3-1}
\end{eqnarray}
Also, we can show that $\mathcal{B}_0'$ connects $\mathcal{B}_6$ 
from the relation $\bm{x}^{inv}_{6}(-\frac{121B}{15}) = (0,\frac{17B}{15})$.
It is found from Fig. \ref{Fig.connection} that the basins in region I are distributed around $(B,-B)$ spirally.
This fact can be confirmed from the following short consideration.
First, the relation 
$\bm{x}^{inv}_{n+1}(z)=\bm A_I^{-1}  \bm{x}^{inv}_{n}(z)+(\frac{B}{2},-B)$
holds for any $z$ by eq.(\ref{eqn:4-4}). 
This relation shows that
$\mathcal{B}_{n+1}$ coincides with the image of $\mathcal{B}_{n}$ by the linear mapping of eq.(\ref{eqn:4-2}).
Since eq.(\ref{eqn:4-2}) is the inverse of eq.(\ref{eqn:2-1}) with $\bm A	= \bm A_I$, 
its dynamics is characterized as the anticlockwise spiral sink with the center $(B,-B)$.
Thus, the basins are distributed spirally.
Note that this property gives rise to the hierarchic and self-similar structure of basins as shown in Fig. \ref{Fig.self-similar}.
%
%

%
%
%
%

%
%

\section{Discussion}
\label{sec:5}

\begin{figure}[h!]
\begin{center}
\includegraphics[width=9cm]{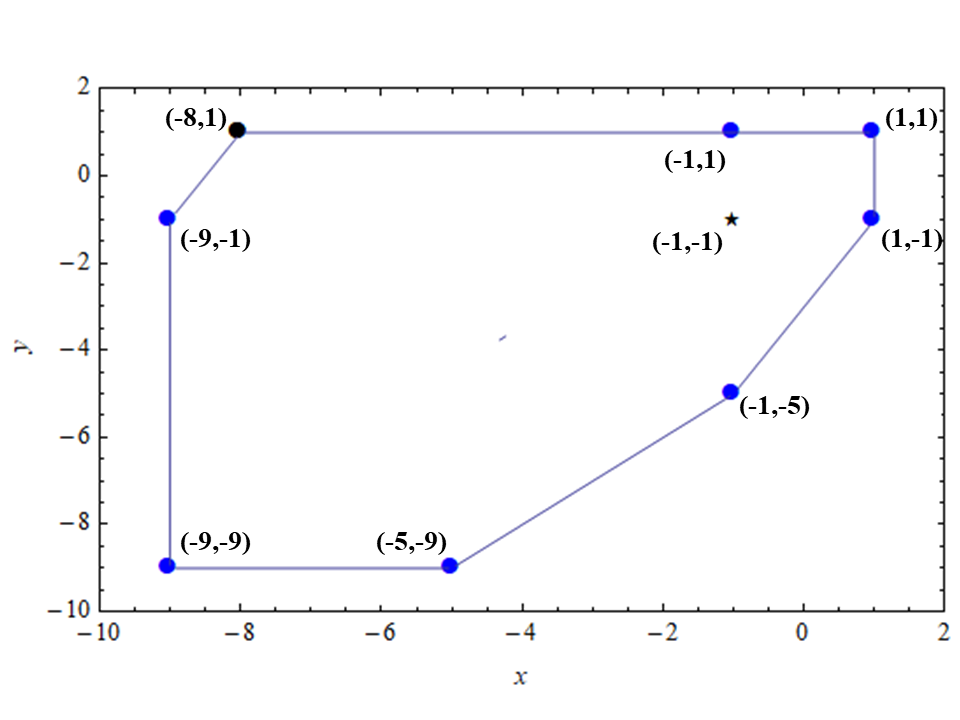}
	\caption{\label{Fig.5-1} 
		The heptagon $\Gamma$ in $xy$-plane.
		The blue solid points provide $\mathcal{C}^\Gamma$.
			}
\end{center}
\end{figure}

Here, we discuss the periodicity of $\mathcal{C}$ and $\mathcal{C}_s$
from a different point of view\cite{Takahashi2021}.
For simplicity, we set $B=1$ in this section.
From eqs.(\ref{eqn:1-1a})-(\ref{eqn:1-1b}),  we have a relation $X_{n+1} + Y_{n+1} = Y_{n} + 1$.
Inserting this relation into eq.(\ref{eqn:1-1b}), we obtain the following piecewise linear mapping with only variable $Y$: 
\begin{eqnarray}
	Y_{n+1} & = & 1 + \min(0, 2(Y_n-Y_{n-1}-1)).
	\label{eqn:5-1}
\end{eqnarray}
Equation (\ref{eqn:5-1}) is composed of two mappings,
$Y_{n+1}=2Y_n-2Y_{n-1}-1$ when $Y_n \leq Y_{n-1}+1$ and $Y_{n+1}=1$ when $Y_n \geq Y_{n-1}+1$.
Here, we consider a heptagon $\Gamma$ 
whose vertexes are given in $xy$-plane ($x=Y_{n-1}, y=Y_n$) as the following seven points, 
$\{ (1, 1)$, $(1, -1)$, $(-1, -5)$, $(-5, -9)$, $(-9, -9)$, $(-9, -1)$, $(-8, 1)\}$.
Figure \ref{Fig.5-1} shows $\Gamma$. 
Note that each $(x,y)$ on $\Gamma$ satisfies the following equation:
\begin{eqnarray}
	\min(y - 2x + 3, y - x + 4, y + 9, x + 9, -y + 2x + 17, -y + 1,-x + 1) = 0.
	\label{eqn:5-2}
\end{eqnarray}
It is noted that the point $(-1,1)$ and the vertexes of $\Gamma$ without $(-8,1)$ compose of 
a clockwise periodic solution $\mathcal{C}^\Gamma$ 
of eq.(\ref{eqn:5-1}),
which is shown in Fig. \ref{Fig.5-1} as the blue filled circles.
$(-8,1)$ is the point mapped by eq.(\ref{eqn:5-1}) from the point $(-9,-8)$, 
which is an intersection between $\Gamma$ and $Y_n = Y_{n-1}+1$. 
%
%
The polygon generated from a piecewise linear mapping, $\Gamma$ in this case,
has been widely studied in the context of ultradiscretization for the integrable systems\cite{Grammaticos1997,Takahashi2002}.
In the following, we apply to $\Gamma$ the method  
used in ultradiscretizing the Quispel-Robert-Thompson system\cite{Takahashi2002}.

\begin{figure}[h!]
\begin{center}
\includegraphics[width=7.5cm]{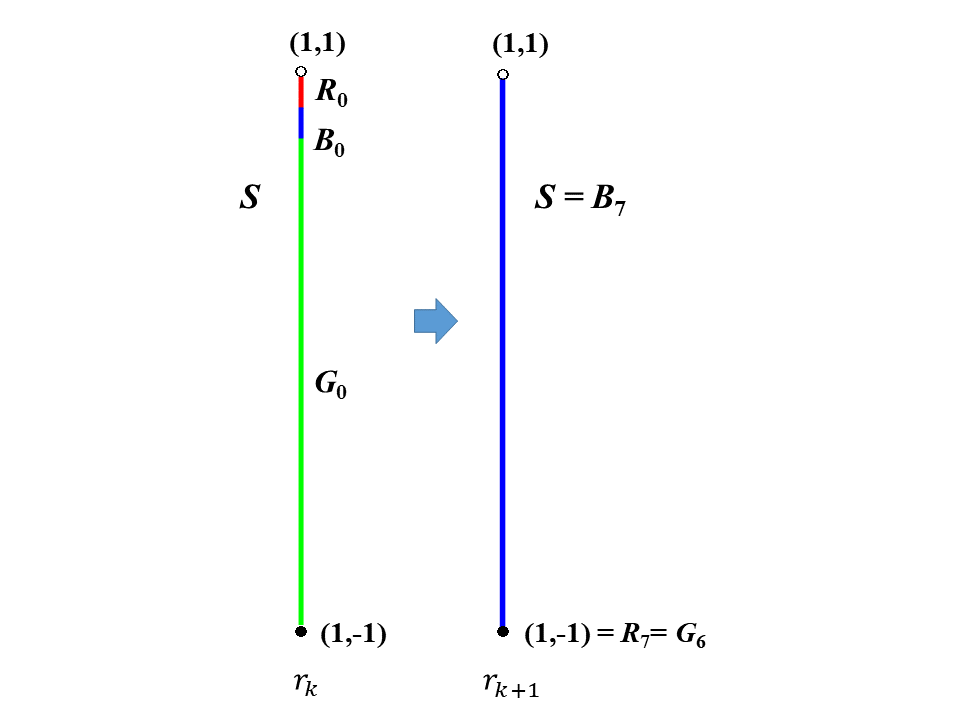}
\includegraphics[width=8cm]{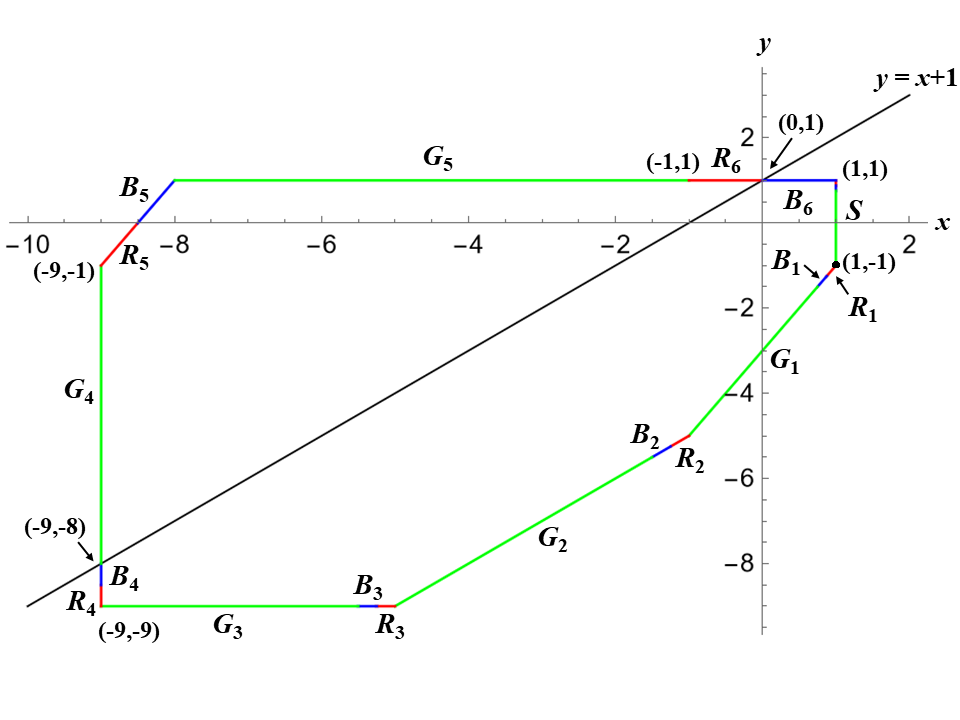}
\\
(a)
\hspace{8cm}
(b)\\
	\caption{\label{Fig.5-2} 
	The sketches of the divisions of (a) $S$ and (b) each segment.
	Equation (\ref{eqn:5-1}) maps the three subsegments $R_0, B_0$, and $G_0$ of $S$ as follows:
	$R_0  \rightarrow  R_1 \rightarrow \cdots \to R_6 	\rightarrow R_7=(1,-1)$,
	$B_0  \rightarrow  B_1 \rightarrow \cdots \to B_6 \rightarrow B_7=S$,
	$G_0  \rightarrow  G_1\rightarrow \cdots \rightarrow G_5 \rightarrow G_6=(1,-1).$ 
	}
\end{center}
\end{figure}

We focus on a half-open line segment from $(1,1)$ to $(1,-1)$ in $\Gamma$, say $S \equiv \{(1,y), -1\leq y <1\}$. 
%
%
All points on $S$ return to $S$ by eq.(\ref{eqn:5-1}) after a certain iteration step.
Here, we divide $S$ into the following three subsegments 
$R_0=\{(x,y) \in S,|(1,1)-(x,y)|\leq \frac{1}{16}\}$, 
$B_0=\{(x,y) \in S, \frac{1}{16} < |(1,1)-(x,y)| < \frac{1}{8}\}$, 
and $G_0=\{(x,y)\in S,\frac{1}{8} \leq |(1,1)-(x,y)|\leq 1\}$ in the ratio of $1:1:14$.
Recall that eq.(\ref{eqn:5-1}) behaves as different linear mappings in the regions $Y_n \leq Y_{n-1}+1$ and $Y_n \geq Y_{n-1}+1$,
and hence the internal ratio for each segment is retained.
By using this fact, each segment composing of $\Gamma$ can be divided into the three subsegments corresponding to $R_0, B_0$, and $G_0$.
Figure \ref{Fig.5-2} (a) and (b) show the divisions of $S$ and of each segment, respectively.
As shown in Fig. \ref{Fig.5-2} (b), the trajectories starting from the three subsegments of $S$ by eq.(\ref{eqn:5-1}) are given as follows:
\begin{eqnarray}
	R_0 & \rightarrow & R_1 \rightarrow \cdots \to R_6 	\rightarrow (1,-1) \nonumber \\
	B_0 & \rightarrow & B_1 \rightarrow \cdots \to B_6 \rightarrow S \nonumber \\
	G_0 & \rightarrow & G_1\rightarrow \cdots \rightarrow G_5 \rightarrow (1,-1). 
	\label{eqn:colors}
\end{eqnarray}
From these trajectories, we find that the only subsegment $B_0$ is stretched over the whole segment $S$ 
whereas $R_0$ and $G_0$ converge to $\mathcal{C}^\Gamma$ after a round of mapping by eq.(\ref{eqn:5-1}).

In order to characterize the self-similar structure of $S$,
let us introduce an internally dividing point $r_k$ defined by
\begin{eqnarray}
	r_{k} & = & \frac{\text{length from $(1,1)$ to $Q_k$}}{\text{length of $S$}} = \frac{|1-q_k|}{2}
	\label{eqn:5-3}
\end{eqnarray}
where $Q_k=(1,q_k)$ 
denotes a point on $S$ at the $k$-th return from an initial point $Q_0=(1,q_0)$ on $S$.
From the trajectories (\ref{eqn:colors}) for $R_0,B_0$, and $G_0$, $r_k$ satisfies the following recurrence formula:
 	\begin{eqnarray}
	r_{k+1} =	\begin{cases}
				1   &  (0<r_k \leq \frac{1}{16}),  \\
				16r_k-1  &  (\frac{1}{16}<r_k < \frac{1}{8}),  \\
				1 & (\frac{1}{8} \leq r_k \leq 1).
			\label{eqn:5-4}
		\end{cases}
	\end{eqnarray}
%
%
$r_k$ has just two fixed points $r_k=1$ and $r_k=\frac{1}{15}$ corresponding to $Q_k=(1,-1)$ and 
$Q_k=(\bar x, \bar y) \equiv (1,\frac{13}{15})$ on $S$, respectively.
Stretching $B_0$ to $S$ is identified with the iteration $r_{k} \to r_{k+1}=16r_k-1$ for $\frac{1}{16}<r_k < \frac{1}{8}$ (Fig. \ref{Fig.5-2} (a)),
and by making use of the graphical analysis as well as Fig. \ref{Fig.P_map}, 
we can confirm that as $k\to \infty$ all point on $S$ without the fixed point $(\bar x, \bar y)$ is absorbed to $\mathcal{C}^\Gamma$.
In other words, only the point $(\bar x, \bar y)$ does not converge to $\mathcal{C}^\Gamma$ but returns to itself by any iteration step of $r_{k} \to r_{k+1}$. 
Therefore, the trajectory starting from $(\bar x, \bar y)$ by eq.(\ref{eqn:5-1}) constructs a cycle with seven points on $\Gamma$, 
say $\mathcal{C}^\Gamma_s$, which is different from $\mathcal{C}^\Gamma$.
Note that $\mathcal{C}^\Gamma$ is stable and $\mathcal{C}^\Gamma_s$ is unstable since $B_0$ is stretched onto $S$.

Equation (\ref{eqn:5-4}) coincides with 
the form of the Poincar\'{e} map (\ref{eqn:3-1-3}) for $B=1$ whose domain is restricted in $(0,1]$.
Actually, the mapping $r_{k}\to r_{k+1}$ can be associated with $P$ as follows.
Let us introduce a mapping $h:(x,y) \mapsto (X_n,Y_n) = (1-y+x,y)$ by which 
$xy$-plane can be transformed homeomorphically onto $X_nY_n$-plane.
Then, eq.(\ref{eqn:5-1}) becomes topologically conjugate to the system of eqs.(\ref{eqn:1-1a})-(\ref{eqn:1-1b}) relative to $h$.
Then, $\mathcal{C}^\Gamma$, $S$, and $\mathcal{C}^\Gamma_s$
transform the limit cycle $\mathcal{C}$, the segment $I$ from $\bm{x}_0^{\mathcal{C}}$ to $\bm{x}_1^{\mathcal{C}}$, and $\mathcal{C}_s$, respectively, 
where $\bm{x}_0^{\mathcal{C}}$ and $\bm{x}_1^{\mathcal{C}}$ are the points composing of $\mathcal{C}$ stated in Table 1 of Sec. 2.
%
%
In $X_nY_n$-plane, we set a point $(x_0,1)$ with $0 < x_0<1$ on $L_B$ ($B=1$).
%
Then $(x_0,1)$ maps to $(1,-2x_0+1)$ on $S$ in $xy$-plane.
Taking $(1,-2x_0+1)$ as the initial point on $S$, 
we obtain from eq.(\ref{eqn:5-3})
	\begin{eqnarray}
		r_0 = \frac{|1-(-2x_0+1)|}{2}=x_0.
			\label{eqn:5-5}
	\end{eqnarray}
When we set $x_0=1$, $(x_0,1)=(1,1)$ on $L_B$ maps to $(1,-1)$ on $S$
and we obtain $r_0=1=x_0$.
Therefore, the value of an internally dividing point $r_0$
is equal to the value $x_0$ on the Poincar\'{e} subsection $(0,1]$ and
the mapping $r_{k}\to r_{k+1}$ completely corresponds to the Poincar\'{e} map $x_{k+1}=P(x_k)$, restricted on $0 < x_k \leq 1$.
%
%
%
Note that the stability of $\mathcal{C}$ and $\mathcal{C}_s$ analyzed by the Poincar\'{e} map $x_{k+1}=P(x_k)$ 
can be also understood through the mapping $r_{k}\to r_{k+1}$ characterizing the self-similar structure of $S$ in $xy$-plane.
We expect that the method developed from the piecewise linear mapping in the integrable system theory 
can be further applied to the study of periodic structures found in generally (non-integrable) max-plus dynamical systems.


\section{Conclusion}
\label{sec:6}

Dynamical properties of the limit cycles $\mathcal{C}$ and 
$\mathcal{C}_s$ found in the max-plus model (eqs.(\ref{eqn:1-1a})-(\ref{eqn:1-1b})) are investigated.
With the aid of Poincar\'{e} map method, 
we confirm their stability; 
$\mathcal{C}$ is stable and $\mathcal{C}_s$ is unstable.
For $\mathcal{C}_s$, we identify its basins 
$\mathcal{B}_n$, $n=0,1,2,\cdots$, and $\mathcal{B}'_0$.
These basins are distributed around the unstable fixed point 
and have a hierarchic and self-similar structure.
We apply the method of piecewise linear mapping studied in ultradiscretizing the Quispel-Robert-Thompson system 
to our max-plus system and reveal the connection of the mapping of the internally dividing point with the Poincar\'{e} map method.

\bigskip

\noindent
{\bf Acknowledgement}

The authors are grateful to 
Prof. D. Takahashi, Prof. T. Yamamoto, and Prof. Emeritus A. Kitada 
at Waseda University for useful comments and encouragements. 
%
This work was
supported by Sumitomo Foundation, Grant Number 200146.

\bigskip



\end{document}